\documentstyle[aps,prbbib,preprint]{revtex}
\begin{document}

\draft
\title{ Relaxation of Thermoremanent Magnetization in Different
 Magnetic Phases Of Fe-rich $\mbox{$\gamma$}$-FeNiCr Alloys}

\author{G.Sinha,$^1$  R.Chatterjee,$^2$\cite{pa}
 M.Uehara,$^2$  and A.K.Majumdar$^1$}

\address{$^1$ Department of Physics,Indian Institute
of Technology, Kanpur 208 016,Uttar Pradesh, India\\
$^2$ National Research Institute for Metals, 2-3-12, Nakameguro,
 Meguro-ku, Tokyo 153, Japan }

\maketitle

\begin{abstract}
The time decay of the thermoremanent magnetization(TRM) of the
$\mbox{Fe}_{80-\mbox{x}}\mbox{Ni}_{\mbox{x}}\mbox{Cr}_{20}$
($14\leq \mbox{x} \leq 30$) alloys has been measured for four
different magnetic phases  within the fcc
$\mbox{$\gamma$}$-phase  using a
SQUID magnetometer. In the spin-glass phase(SG)(X=19) very distinct
ageing effects are observed where M(t) can be described
as $M(t)=M_0(t/t_{w})^{-\gamma}\exp[-(t/\tau)^{1-n}]$ for the
entire time domain.  In
the reentrant spin-glass(RSG)(X= 23 and 26), M(t) can be better
represented by
the stretched exponential  with an addition of a constant term
which can be well explained by the Gabay-Toulouse(GT) model.
We have also measured the linear and non-linear ac susceptibilities
for the sample X=23 and confirmed the presence of the
ferromagnetic(FM) ordering down to the lowest temperature. In
the RSG(X=23), the TRM shows a minimum near T$_{\mbox{c}}$ and a
local maximum just above T$_{\mbox{c}}$.
In the FM phase (X=30)
 the popular prediction of the power law decay of the TRM is observed.
  The latter  is
indistinguishable from the  stretched
exponential in the antiferromagnetic(AF) phase (X=14).

\end{abstract}


\section{INTRODUCTION}

Since the last decade,
 $\mbox{Fe}_{80-\mbox{x}}\mbox{Ni}_{\mbox{x}}\mbox{Cr}_{20}$
 ($14\leq x \leq 30)$
alloys have been the subject of considerable interest because of
their diverse magnetic properties within the same
crystallographic phase.\cite{akm} These alloys exhibit a
compositional phase transition from a long-range AF phase(X=14)
to a long-range FM one(X=30), passing through intermediate
phases of SG(X=19) and RSG(X=23,26) with the increase of Ni
concentration.  The presence of the strong competing ferro-
[I(Ni-Ni), I(Fe-Ni), I(Ni-Cr), I(Fe-Cr)] and
antiferromagnetic[I(Fe-Fe), I(Cr-Cr)] exchange interactions
\cite{men's} is responsible for  such rich magnetic
phases. The complete phase diagram of these alloys was
established by Majumdar and Blanckenhagen through dc
magnetization and neutron diffraction measurements.\cite{akm}
Later on ac susceptibility \cite{sbroy} and  magnetoresistance
measurements \cite{tkn} also confirmed the  proposed phase
diagram. Taking advantage of the  diverse magnetic properties of
these alloys {\em within the same fcc $\mbox{$\gamma$}$-phase},
we  tried to understand the age-old problem of relaxation
dynamics. This area has  always been in the front-line because
of its  multiple complexities and wide varieties.
In this paper, we  present the dynamics of time and
 temperature-dependent magnetic relaxations  in various magnetic phases.

The search for magnetic relaxation began a century back, when
Ewing \cite{ewing} observed   persistence of magnetization in
soft iron  for significant amount of time and  a non-exponential
decay.    Richter \cite{richter} observed logarithmic
decay of magnetization for about one decade of time in
carbonised iron.   Street and Woolley
\cite{sw} and Neel\cite{neel} also predicted a logarithmic decay
of TRM in FM.
 Several theoretical and experimental evidences suggest
 that an anomalous slower relaxation of the form
\begin{equation}
M(t)=M_0\exp\,[-(t/\tau)^{\beta}]\;,\;
\;0<\beta<1\;,
\end{equation}
 is far more reasonable and common than the
conventional Debye exponential form($\mbox{$\beta$} =1$). In
fact, this kind of relaxation had been observed for a wide range
of phenomena and materials.\cite{ngai}  In 1970, Williams and
Watts
\cite{ww} postulated similar functions for dielectric relaxation
($\mbox{$\beta$}$ = 0.5). In a review article Jonscher
\cite{jonscher}  summarized the experimental evidence on
the frequency, time, and temperature dependence of the
dielectric response for a wide range of solids.  He also found a
universality of dielectric behaviour and proposed a generalized
approach of many-body interaction.  The structural relaxation
rate, in the case of liquid to glass transition, can be expressed as
a stretched exponential  (2/3 $<$ $\mbox{$\beta$}$ $<$ 1).
\cite{cohen}  Also the validity of this functional form for the
relaxation of TRM in SG has been reported by a large number of
investigators.\cite{rvc84,rhoo,fmezei,rvc85,nordblad}
Palmar et al. \cite{pal} presented in an elegant fashion the
whole scenario of similar kinds of relaxation in complex, slowly
relaxing, and strongly interacting materials.
They  considered series relaxation and hierarchically constrained
dynamics which is distinctly different from other approaches
\cite{mfs} of getting similar results.
 Hammann et al. proposed a phenomenological picture of the
dynamic properties  of  SG based on fractal cluster
model.\cite{ham,ll}  Early  decay measurements of  TRM of
spin-glasses  had shown logarithmic dependence.\cite{rt}
Similar dependence was also observed in
$\underline{\mbox{Au}}$Fe, $\underline{\mbox{Ag}}$Mn,
$\underline{\mbox{Th}}$Gd  spin-glasses from  5 to 10$^4$
s.\cite{so} Analysing the neutron diffraction and ac
susceptibility data,  Murani \cite{apm} found power law decay
for shorter time and logarithmic decay for longer time below
T$_{\mbox{g}}$ in SG. Bontemps and Orbach\cite{nbon} measured
the TRM of
 the insulating SG Eu$_{0.4}$Sr$_{0.6}$S between 0.86T$_g$ and
 1.04T$_g$. They found power law decay of the TRM for shorter
 interval of time and a stretched exponential decay beyond
 a well-defined cross-over time.
Recently two different theoretical models
have been proposed to describe the SG behaviour.
The first is the mean-field
approach of Parisis's solution \cite{parisi} of infinite range
Sherrington-Kirkpatrick(S-K) \cite{sk} model by considering
hierarchical organizations of infinite number of
quasi-equilibrium states in phase space.\cite{sibani}
The second  is the
phenomenological approach based on the existence of a
distribution of droplets \cite{fisher} or dynamical domains
\cite{koper} of correlated spins.
Both these
theories can explain reasonably well  the slower SG dynamics and
ageing effects.  A nice comparison between  these two models was
 given by Lefloch et al.\cite{lef}  Huse and Fisher
\cite{fisher} proposed long time decay as stretched exponential with
exponent 1/2 by using droplet fluctuations in two dimensional
pure Ising system with a spontaneously broken continuous
symmetry.  In the framework of droplet
fluctuations,\cite{fisher} the spin autocorrelation can be described
as exponentially rare in $\ln t$ :
$\overline{\mbox{c}_{\mbox{i}}(\mbox{t})}$ = $\exp[-k(\ln
t)^{y}]$ (where $y = 1$ for random exchanges, $y = d-2/d-1$ for
random fields) in random FM and power of $\ln t$ in SG.
 Ogielski\cite{ogi} predicted that the spin autocorrelation
 function can be described by the product of power law and
 stretched exponential at all temperatures above T$_g$\cite{gunn}
   and by power law below T$_g$.
  Ocio et al. found an analytical expression based on scaling
  analysis of ageing effects, for the decay of TRM in SG,
  CsNiFeF$_6$ \cite{ocio} as well as in  AgMn.\cite{alba}
They proposed the decay of TRM as
\begin{equation}
M(t)=M_0\;(\lambda)^{-\alpha}\,\exp\,[-(\omega\, t/t_{w}^{\mu})^
{1-n}]\;,\;
\end{equation}
\begin{equation}
 with\,\,\;\lambda=(t_{w}/(1-\mu))[(1+t/t_{w})^{1-\mu}-1]
\;,
\end{equation}
where $\mbox{$\lambda$}$ is the effective time and
$\mbox{$\mu$}$  is an exponent smaller than 1 .
 A different  analysis, based on S-K mean-field model,
 \cite{hs}  suggests algebraic decay.
 In ferromagnets attempts have been made to
explain the magnetic relaxation using a model based on magnon
relaxation on a percolation distribution of finite domains.
\cite{rvc90}  The other popular prediction  of relaxation is the
power law \cite{rjb} decay which can be obtained from scaling
theories for domain growth \cite{gunton} and internal
dynamics.\cite{halp}  Ikeda and Kikuta\cite{ike} found that there is
 no magnetic relaxation in AF over 10 h in $Mn_{0.45}Zn_{0.55}F_2$
  at a slightly lower temperature than the transition
 temperature.

Despite considerable experimental work on relaxation dynamics
covering enormous range of time window, no conclusive results
have been found. Moreover, there is a lack of  clear distinction
between the SG and the RSG phases. Also no  experimental
data are  available on relaxation dynamics in the AF phase. All these
have motivated us  to study systematically  the relaxation
dynamics in four different magnetic phases, namely,  SG, RSG, FM
and AF in FeNiCr alloys {\em within the same crystallographic
phase}, for different wait times and at different temperatures
for the largest available time window. There is a running
controversy about the existence of the FM orderings in RSG
at the lowest temperature as predicted by the GT model.\cite{gt}
We have also measured
linear and non-linear ac susceptibilities for the sample X=23
to resolve this controversy.

\section{EXPERIMENTAL PROCEDURE}

All these ternary alloys were prepared by induction melting in
argon atmosphere. The starting materials were of 99.999\% purity
obtained from M/s Johnson Mathey Inc., England. The alloys were
cut to the required size, homogenized at 1323 K for 100 h in
argon atmosphere, and then quenched in oil.

Chemical analysis of Ni and Cr shows that the compositions of
the alloys are within $\pm 0.5$ at. $\%$ of their nominal values.
X-ray diffraction data at room temperature in powdered  samples
reveal that these  are single-phase fcc ($\gamma $) alloys with the
lattice parameter a=3.60 $A^{o}$. Neutron diffraction data show
the presence of single-phase fcc ($\gamma $)structure
 down to 2 K for X=19
alloy.

The two most essential ingredients of SG are "frustration" and
"quenched disorder" which manifest themselves  in history
dependent phenomena and eventually lead to nonequilibrium
behaviour below T$_{\mbox{g}}$.\cite{binder} To study the
relaxation dynamics, we applied a small magnetic field (10
gauss) at a temperature greater than the characteristic
temperatures (T$_{\mbox{g}}$ , T$_{\mbox{c}}$ , T$_N$ ) , cool
($\mbox{t}_{\mbox{c}}$ = cooling time ) the system to a
temperature T$_{\mbox{m}}$  which is  less than the
characteristic temperature.  This field-cooled system will
attain a metastable state (which is not an equilibrium state),
\cite{rhoo,rvcprb84}  then after waiting  for  definite amounts of time
$\mbox{t}_{\mbox{w}}$ ( $\mbox{t}_{\mbox{w}}$ varies from 60 to
3600 s )  the magnetic field is removed,  the system allowed to
relax towards equilibrium state and then TRM measured with time
(till $\sim 10^{4}$ s) using a SQUID magnetometer
(Quantum Design MPMS).  To see the
effect of temperature, we also measured TRM at several
temperatures, T$_{\mbox{m}}$,  both above and below the
characteristic temperatures for constant wait time,
t$_{\mbox{w}}$ = 180 s.  We have very carefully subtracted  the
background noise. We repeated the experiments  and obtained
similar  results within the experimental error.

\section{RESULTS AND DISCUSSION}

We observe  remarkable results of ageing effects in the SG where
each isotherm strongly depends upon wait time t$_{\mbox{w}}$
(time of exposure in magnetic field below T$_{\mbox{g}}$). The
magnetization  can be well represented by an equation of the
form
\begin{equation}
M(t)=M_0(t/t_{w})^{-\gamma}\exp\,[-(t/\tau)^{1-n}]
\end{equation}
for the entire time domain.
This is a simpler version of the earlier prediction of
scaling analysis of ageing process \cite{alba,ocio} (Eq. (2)).
Instead of t they used $\lambda$ which is a function t and t$_w$
 and $\lambda$ $\rightarrow$ t for t $<<$ t$_w$. But we find
 that the Eq. (4) is valid for the entire time domain.
 Ogielski\cite{ogi} predicted similar analytical form to
 describe the spin autocorrelation function in SG at all
 temperatures above T$_g$.
 However other investigators
\cite{rvc84,rvc85,nordblad,chu} found only  the stretched exponential form for
the decay of TRM for the  SG phase.  We observe in the RSG the
best representation of TRM  is
\begin{equation}
M(t)=M_{1}+ M_0\,\exp\,[-(t/\tau)^{1-n}]\,\;.
\end{equation}
The additional small term $\mbox{M}_1$ can easily be explained
in the framework of the GT model \cite{gt} where only transverse
spin freezing occurs in the RSG while the longitudinal spins can
produce diffuse background effect. Similar expression also
reported for the decay of TRM in SG as well as in RSG.\cite{p-m}
 They did not find any difference between the RSG and the SG
 phases. We find that  Eq. (5) is only valid for the RSG and
 distinct differences exist between the SG and the RSG phases.
The most salient feature  of the
present work is that it can demonstrate how  changing the
composition by small amounts  in
$\mbox{Fe}_{80-\mbox{x}}\mbox{Ni}_{\mbox{x}}\mbox{Cr}_{20}$
alloys ( X = 14 AF, X = 19 SG, X = 23,26 RSG, X = 30 FM ) one
can tune the relaxation dynamics. This is the only experimental
report which presents a complete scenario of the relaxation
spectrum in various kinds of interesting magnetic phases and
throws new light on this area  of interest to theoreticians as
well as  experimentalists .

Figure 1 shows the time decay of TRM, M(t), for different wait
times (t$_{\mbox{w}}$ = 60, 240, 1200, 1800, 3600) below
T$_{\mbox{g}}$(12 K) (T$_{\mbox{m}}$ = 5 K) for the SG (X = 19) and
the solid lines are the best fits of the experimental data
to Eq.(4) for almost four decades of time (till  12,000 s)
($\chi^2=(1/n)$\(\sum_{i=1}^{n}(Raw\;data_i-Fitted\;data_i)^2/Raw\;data_i^2\)$
\leq 10^{-6}$).
We have purposefully plotted our data on linear time scale
because if we plot on log scale, for longer time interval, the
plot will contract and the fits will apprently look better.
However, goodness of fit is better judged from the value of
$\chi^2$.
 From these fits, the value of the initial TRM, $\mbox{M}_0$ ,
the characteristic time constant, $\mbox{$\tau$}$, and the
exponents, n and $\mbox{$\gamma$}$,  are found. The most
significant feature of this analysis is that M$_0$ ($\approx$
0.04 emu/g)  is not varying with t$_{\mbox{w}}$
\cite{rvc84,rhoo,alba} while the exponent, n,  gradually
increases with the increase of t$_{\mbox{w}}$. This indicates
that larger time exposure in magnetic field below T$_{\mbox{g}}$
makes the system  more reluctant to come back to an equilibrium
state from a metastable one. Other investigators
\cite{rvc84,rhoo} reported n independent of wait time but we
found that it varies from 0.63 to 0.77.  The power law exponent,
$\mbox{$\gamma$}$, remains constant (0.022 $\pm$ 0.004) except
$\mbox{t}_{\mbox{w}}$ = 60 s where $\mbox{$\gamma$}$ is 0.03.
Figure 2 shows the variation of TRM , M (t), with time for
different temperatures at constant wait time
($\mbox{t}_{\mbox{w}}$ =180 s). From the best fit to Eq.(4) the
temperature variations of n, $\mbox{M}_0$, $\mbox{$\tau$}$,  and
$\mbox{$\gamma$}$ are found which broadly match  with the
previous observations.\cite{rvc84,rhoo,alba,chu}  The value of
n  increases linearly from 0.8  to  0.9  from 0.5T$_{\mbox{g}}$
to  T$_{\mbox{g}}$ and then it starts  falling beyond
T$_{\mbox{g}}$ (Fig. 3). It is reported that  n remains constant
at lower temperatures and then  starts  rising from  a
temperature T =T$_0$, which  strongly depends on the anisotropy
energy of the sample \cite{rvc85} and hence can vary from sample
to sample.  Non-availability of lower temperature data(less than
0.4T$_{\mbox{g}}$)  prevented us from verifying the constancy of
n.  The  prefactor, M$_0$, shows (Fig. 3) a linear decrease with
increasing temperature for T $<$ 0.75 T$_{\mbox{g}}$ and the
rate of decrease becomes slower for T $>$ 0.75 T$_{\mbox{g}}$,
which is in good agreement with the earlier findings.\cite{rvc84} The
power law  exponent $\mbox{$\gamma$}$ increases with the
increase of temperature up to 0.75 T$_{\mbox{g}}$. At
T$_{\mbox{g}}$ and beyond  it starts decreasing with temperature
(Fig.~3). It is difficult to tell  exactly  the  temperature from
which it
has started falling , but  earlier prediction was that it should
increase as one approaches T$_{\mbox{g}}$.\cite{alba} Figure 4
shows how the inverse of the characteristic relaxation time
$\mbox{$\tau$}$  decreases with the increase of the reduced
temperature T$_{\mbox{g}}$/T,  as observed by Alba et al.\cite{alba}  It also
reflects the general tendency  observed by
others.\cite{rhoo,chu} Fewer number of data points prevents us
from finding the functional form. Apparently, to check the
exponential form as predicted earlier\cite{rhoo} we need to probe even
at lower temperatures. We can try to  explain the variation of
$\mbox{$\tau$}$ with temperature by considering hierarchically
organized metastable states in phase space \cite{lef} within the
framework of Parisi's  mean-field solution of infinite range S-K
model.  The  states are continuously splitting into new states
with the lowering of temperature and are separated by barriers.
These barrier heights strongly vary inversely with temperature.
That is, at lower temperatures, barrier heights increase and
separate different metastable states into mutually inaccessible
states  which makes $\mbox{$\tau$}$ larger at lower
temperatures.  Using values of n and $\mbox{$\tau$}$ from Figs.
3 and 4, respectively and plotting $\log(1-n)(1/\tau)^{1-n}$
versus (1-n) by writing this function as $(1-n)(1/\tau)^{1-n}$ =
$c\,
\omega^{1-n}$, we get the value  of the relaxation frequency
$\mbox{$\omega$}$ from the slope of the best fitted curve
(Fig.5) as 1.9$\times10^{-7}$ s$^{-1}$  and the constant, c, is
0.13.  These values are  much less than those predicted
earlier.\cite{rvc84}
  Analysis of M(t) by considering only the stretched
exponential,  ie., without the power law part of Eq.(4),  shows
unusually small ($\approx 10 $ times smaller) values of n  and a
poor $\mbox{$\chi$}^2$($\approx 10^{-5}$).  Fits are even poorer in cases of
other
mathematical forms,  like the power law.  So, we find  that
Eq.(4) is the simplest analytical form that can represents our
experimental data of the decay of TRM in the
SG phase for the entire time domain.

 The SG has been the focus of attention for quite
sometime, but not  much attention has  been paid to the RSG phase.
The sample with  X = 23, below 35 K,  enters   a  FM phase from
a random paramagnetic(PM) phase. On further lowering of temperature
below 22 K it enters once again  a new random phase where FM and
SG orderings coexist.\cite{akm}  It shows most of the SG-like
behaviour (frustration, irreversibility, etc. ).  This phase is
known as the RSG.  Figure 6 represents the variation of M(t)
with time at T$_m$=5 K for different wait times (t$_{\mbox{w}}$ = 60, 1200,
1800, and 3600 s ) in the RSG phase (X=23) and the solid lines
are the best fitted curves which are of the form of Eq.(5)($\chi^2
\approx 10^{-6}-10^{-7}$).  The
salient feature of this figure is that the initial
magnetization, $\mbox{$\sigma$}_0$ = M$_0$ + M$_1$, increases
with the increase of wait time (Fig. 7). This  is not observed
in the SG phase but the variation of n  is similar.  We also
observe that M$_1$, which is arising because of the presence of
the FM ordering,
\cite{gt} is not changing at all with wait time and the   values
of M$_1$ are 84$\%$ to 94$\%$ of the total magnetization
depending on wait time (total magnetization changing with wait
time).  It is quite natural that if the ferromagnetic component
is embedded in the SG phase, then the major contribution of the
total magnetization should come from the ferromagnetic
component.  Figure 8 depicts the variation of M(t) with time at
different temperatures ( T$_{\mbox{m}}$ = 10, 15, 20, 25, 30, 38
K
) for constant wait time (180 s). From the fits of the data to
Eq. (5), the value of
$\mbox{$\sigma$}_0$ and the exponent, n, are found. The
exponent increases with temperature and approaches  unity as the
temperature approaches T$_{\mbox{g}}$. This is exactly in
agreement with the previous observations \cite{rvc84,rhoo} in
SG,  the only difference is that the values are slightly higher.
If we increase the temperature even more, a sudden change takes
place around 20~K.  A drop in the value of  n  indicates
 a phase change.   Moreover, at lower temperatures it
shows better fit to Eq.~(5) but at higher temperatures  power
law fit is better than the stretched exponential.  This
supports the on-set of the  FM phase.
This could also be an indication of the switch-over from nonequilibrium
dynamics to
equilibrium dynamics  when it passes from the RSG to the FM
phase.
The power law behaviour in
the FM phase (having an exponent  $\approx$ 0.06) is quite
consistent with the Huse and Fisher theory.\cite{fisher}
 The rate of increase of the exponent also changes somewhat
beyond 30 K where it passes from the FM to the PM state at 35 K.
 The value of the exponent varies by 4.6 $\%$ in the temperature
 interval of 8 K (30 K to 38 K)(Fig. 9). This is above the error
 bar which is less than 1 $\%$. The size of the symbol (Fig. 9) is of
 the order
 of the error bar. The value of the
exponent shows anomaly near two temperatures,  22 K and 35
K, which are nothing but T$_{\mbox{g}}$ and T$_{\mbox{c}}$,
respectively.\cite{akm} But the variation of n near T$_c$ is not
as prominent as that of near T$_g$. We also find that in Eq.~(5) the
additional term , M$_1$,  which is the value of the residual
magnetization, M($\infty$),  decreases with the increase of
temperature below T$_{\mbox{g}}$ (0.0067 emu/g at
T$_{\mbox{g}}$) and suddenly increases to 0.027 emu/g when it
enters the FM phase, as expected.  The most interesting
observation is the variation of the initial magnetization,
$\sigma_0$, with temperature (Fig. 9).  $\mbox{$\sigma$}_0$
decreases monotonically with the increase of  temperature up to
20 K beyond which the rate of decrease reduces significantly and
at 25 K and 30 K it becomes almost constant (0.033 and 0.032
emu/g,  respectively) and then there is a sudden rise at 38 K (0.045
emu/g)
which makes the scenario most interesting. The value of $\sigma_0$
changes about 40$\%$ in this temperature interval (30 to 38 K).
This kind of
remarkable observation of switching of magnetization while
passing from FM to PM phase was reported earlier only by
Chamberlin and Holtzberg \cite{rvc91} in ferromagnetic  EuS single crystal.
They tried to explain this in terms of the percolation
theory.\cite{stauffer} So we observe that the
TRM in RSG (X=23) shows a minimum near T$_c$ and a local maximum just
beyond T$_{\mbox{c}}$.
 At lower temperatures
($<$ T$_{\mbox{c}}$), the finite size domains try to orient
themselves with the direction of the local field, which need not
be in the same direction as the applied field. These domains are
dynamically strongly correlated (forming a strong viscous
medium). FeNiCr alloys have shown very large high-field
susceptibility,\cite{mit} i.e., even at very  high magnetic fields
the
orientation of these domains with the direction of the applied field
is not complete because of the presence of strong anisotropy.
Near T$_{\mbox{c}}$ the correlation between finite domains gets
disrupted. Just above T$_{\mbox{c}}$,  the domain magnetization
still remains but the domains become less viscous and they try
to orient themselves along the direction of the applied field,
thus increasing the magnetization. Further increase of
temperature randomizes the whole spin orientation.  We repeated
the experiment  to confirm this unusual observation and found
 similar  behaviour.

In general the non-linear magnetization in the presence of a
magnetic field(h), can be written as $ m = m_{0}+\chi_{0}h +
\chi_{1}h^2 + \chi_{2}h^3 + ....$, where $m_0$=spontaneous
magnetisation, $\chi_0$ linear- and $\chi_{1},\chi_{2}, etc.,$
are nonlinear susceptibilities, and $h=h_0\; Sin\; \omega t$ for
a.c field. In SG where no spontaneous magnetization exists, only
odd harmonics of the susceptibility will  be present.
\cite{tosi,masuo}  We observe  that the out-of-phase part of the
linear a.c susceptibility ($\chi_{0}^{\prime\prime}$) shows
peaks
\cite{masuo} near T$_{\mbox{g}}$ and T$_{\mbox{c}}$,
respectively (Fig. 10) (a.c field of 0.6 Oe and frequency  $\omega$
=242 Hz).  Similar peaks are observed in
$\chi_{2}^{\prime\prime}$ ($\partial^3m/\partial
h^3$, out-of-phase part of the 3rd harmonics).  The distinct peak
near T$_{\mbox{g}}$ for X=23 is consistent with the
theoretical predictions.\cite{masuo,e-a}  $\chi_{1}$
($\partial^2m/\partial h^2$,  2nd harmonics)
also shows a distinct peak near T$_{\mbox{g}}$ (Fig. 10) which is
never observed  in pure SG, including X=19. This indicate the
presence of a ferromagnetic component in RSG down to the lowest
temperature.  So, we find that the RSG shows the SG transition
and also the presence of  FM ordering below T$_{\mbox{g}}$.
This  distinguishes  it from the SG. Detailed studies of ac
susceptibilities in FeNiCr alloys will be published elsewhere.

The sample with X=26, below 56 K , enters a FM phase from a
random PM phase. On further lowering of temperature below 7 K it
enters a RSG phase.\cite{akm}  We observe some  different
features in  the samples X=23 and X=26, though they
undergoes similar kinds of phase transitions(PM $\rightarrow $ FM
$\rightarrow$ RSG).  M(t) data at 5 K, similar to those shown in
Fig. (6) but now for the sample X=26, when fitted to Eq. (5)
($\chi^2\approx 10^{-6}-10 ^{-7}$),
shows that the initial magnetization, $\mbox{$\sigma$}_0$ =
M$_0$ + M$_1$, does not change with wait time ($\approx 0.5 $
emu/g). The value of the exponent, n, increases from 0.55 to
0.64, with the increase of wait time from 240 s to 3780 s. We
have observed similar variation in the SG(X=19) phase whereas in
the other sample, X=23, $\mbox{$\sigma$}_0$ increases with wait
time in the RSG phase. The value of $\mbox{$\sigma$}_0$ is
larger for the sample X=26 than that for the sample X=23 and
smaller than that   for the sample X=30(FM, describe below).
Hence the value of the initial magnetization,
$\mbox{$\sigma$}_0$, increases gradually with the increase of Ni
concentration as we move towards the FM phase.  This is exactly in
agreement with the previous observations.\cite{akm} Variation of
n is similar to both the SG(X=19) and the RSG(X=23) samples.  Figure 11
displays the variation of M(t) with time at different
temperatures (T$_{\mbox{m}}$ = 6, 8, 10, 20, 30, 40, 50, and 60
K) for constant wait time (180 s)for the sample X=26. From the fits
($\chi^2\approx 10^{-6}-10 ^{-7}$)
of the data to Eq. (5), the values of
initial magnetization, $\mbox{$\sigma$}_0$, and the exponent, n,
are found.  $\mbox{$\sigma$}_0$ decreases at a faster rate up
to 20 K and then continues to decrease slowly till 60 K (Fig.
12). We have not observed the local maxima above T$_{\mbox{c}}$
as in the case of X=23. To observe this we need to probe at a
temperature closer to  T$_{\mbox{c}}$(56 K).  The exponent, n,
increases abruptly when the system undergoes a phase transition
from the RSG to the FM at 7 K. Then it starts to  decrease up to 20
K and beyond this again increases till 50 K. Further increase of
temperature reduce the value of n and the system passes from the
FM to the PM phase (Fig. 12). The variation of n is not well
understood, specially the  dip around 20 K.
Interestingly, we have also observed around this temperature some striking
features
 in the low-field magnetoresistance  and
a.c susceptibility measurements.\cite{tbp}
These have  kept  the field wide open for further work. We also observe
that at higher temperature (T$\geq$30 K, much higher than T$_g$)
M(t) data, show better fits to the power law compared to
the stretched exponential (Eq. (5)). For X=23, we observe similar
behaviour for T $\geq$ T$_g$.

Figure 13 shows the variation of  TRM with time at 5 K in the FM
phase (X=30) for different wait times (t$_{\mbox{w}}$ = 240,
1380, 1980, 3780 s). We find that
\begin{equation}
M(t) = M_1 +
M_0\;t^{-\gamma}
\end{equation}
is the best fit of the
experimental data (solid lines in the graph) and from the fits
($\chi^2\approx 10^{-6}-10 ^{-7}$)
the values of $M_1$,  $M_0$ and  $\gamma$ are found.  It does
not show much wait-time dependence,  the values of $M_1$  and
$M_0$ remain almost constant ($\approx$ 0.29 and 0.35 $\pm$
0.008 emu/g, respectively) while the exponent, $\gamma$,
decreases inversely with wait time (from 0.049 to 0.036). With
the increase of temperature the value of the initial
magnetization reduces drastically (0.65 emu/g at 5 K and 0.094
emu/g at 80 K) and the exponent also becomes  smaller (0.049 at
5 K and 0.00036 at 80 K).

Figure 14  shows the time decay of TRM, M(t), for different wait
times (t$_{\mbox{w}}$ = 180, 1320, 1980, 3780 s) below
T$_N$=26~K (T$_{\mbox{m}}$ = 5 K) for the AF(X = 14) sample and
the solid lines are the power law  fits of the form
\begin{equation}
M(t) = M_0
t^{-\gamma}.
\end{equation}
 From these fits the value of $M_0$ and
$\gamma$ are obtained (Table I).
 $M_0$ ($\approx$ 0.00206 emu/g)
does not change with wait time  whereas $\gamma$
decreases with the increase of wait time (0.0044 to 0.0038).
Figure 15 shows the time decay of TRM, M(t), at different
temperatures for constant wait time (180 s) and the solid lines
are the power law fits from which the value of $M_0$ and
$\gamma$ are found (Table III). The value of $M_0$ decreases
linearly with the increase of temperature up to T$_N$ (0.0015
emu/g at 24 K) and then suddenly falls  to a much lower value
(0.00012 emu/g at 30 K) as shown in Fig. 16. The exponent,
$\gamma$, increases with  temperature
and the rate of increase changes somewhat when it passes to the PM phase
(Fig. 16). In case of the AF phase we need not add any constant
term (unlike  the FM phase) and the value of the exponent is an
order of magnitude lower than that  in the FM phase. We also
observe that the stretched exponential function, Eq.(1), shows
reasonably good fits to the TRM in the AF phase. The value of
the fitted parameters and the $\chi^2$ are given in Table II and
IV.
The value of the exponent, $\mbox{$\beta$}$, is almost two orders of
magnitude smaller than that  in the SG phase (0.004 and 0.37,
respectively). It increases monotonically (Table IV) with temperature
(0.004 at 5 K and 0.029 at 30 K) in contrast to that in the  SG
phase where  $\mbox{$\beta$}$ decreases with the increase of
temperature up to T$_
{\mbox{g}}$ (0.37 at 5 K and 0.1 at 12 K obtained from Fig. 3
using $\beta$ = 1-n). So, we find that the M(t) data for AF fit
well to both the power law (Eq. (7)) and the stretched exponential
 (Eq. (1)).
We have given the values of $\chi^2$ in Table I-IV for
comparison. They  are comparable for both the
above mathematical forms. Hence it is difficult to describe the
exact nature of the decay of the TRM in the AF phase. More
experimental work is  needed to arrive at a more definite
 conclusion.

\section{conclusion}

We have measured the TRM in
$\mbox{Fe}_{80-\mbox{x}}\mbox{Ni}_{\mbox{x}}\mbox{Cr}_{20}$
($14\leq \mbox{x} \leq 30$) alloys
  for four different magnetic phases
within the {\em {same crystallographic phase}} and from their
 wait time and temperature variations   tried to establish a
 correspondence with the phase diagram.
We find  distinct differences between the SG and the RSG phases
 with  two different analytical forms for the time decay of
 the TRM.
 We also observe the presence of the FM ordering in the RSG
below T$_{\mbox{g}}$ which is consistent with the GT model.
 The peak in $\chi_1$  near T$_{\mbox{g}}$ for RSG (X=23) had never
been observed in any
pure SG, including X=19. The peak can only be observed if spontaneous
magnetization is present. Hence we confirm the presence of the FM
ordering
 in RSG down to the lowest temperature. This feature distinguishes the  RSG
from the SG.
 We also report the remarkable observation of the local maximum of
 the
TRM just above T$_{\mbox{c}}$ in the RSG (X=23) when it passes to
the PM phase from the FM phase. This is found for the first time
 in any polycrystalline RSG. However, it's exact theoretical justification
 is unclear. We also observe that the value of the exponents
 show anomaly near the phase transitions.
 We observe some different features in the samples X=23 and X=26,
though they undergoes similar kinds of phase transitions.
 We find  the
conventional power law decay in the FM  phase.
 The value of the magnetization is found to increase with the Ni
concentration, as predicted earlier.  In the AF phase the power law  decay
is indistinguishable from the  stretched
exponential as a  description of the TRM. More experimental work is
 needed to arrive at a more definite  conclusion about the decay of TRM in AF.

\section{Acknowledgement}

Financial assistance from project No. SP/S2/M-45/89 of the
Department of Science and Technology, Government of India, is
gratefully acknowledged.

\newpage
\end{document}